\documentclass[conference]{IEEEtran}
\IEEEoverridecommandlockouts
\usepackage{cite}
\usepackage{booktabs}
\usepackage{amsmath,amssymb,amsfonts}
\usepackage{algorithmic}
\usepackage{graphicx}
\usepackage{textcomp}
\usepackage{xcolor}
\usepackage{hyperref}
\usepackage{fancyhdr} 
\usepackage{afterpage} 

\def\BibTeX{{\rm B\kern-.05em{\sc i\kern-.025em b}\kern-.08em
    T\kern-.1667em\lower.7ex\hbox{E}\kern-.125emX}}

\fancypagestyle{firstpage}{
  \fancyhf{} 
  \fancyhead[L]{2023 5th International Conference on Telecommunication and Photonics (ICTP)\\December 21-23, Dhaka, Bangladesh} 
  \fancyfoot[L]{979-8-3503-9347-7/23/\$31.00 ©2023 IEEE} 
}

\begin{document}

\title{Obfuscated Malware Detection: Investigating Real-world Scenarios through Memory Analysis\\
}

\author{\IEEEauthorblockN{S M Rakib Hasan}
\IEEEauthorblockA{\textit{Department of Computer Science and Engineering} \\
\textit{BRAC University}\\
Dhaka, Bangladesh \\
sm.rakib.hasan@g.bracu.ac.bd}
\and
\IEEEauthorblockN{Aakar Dhakal}
\IEEEauthorblockA{\textit{Department of Computer Science and Engineering} \\
\textit{BRAC University}\\
Dhaka, Bangladesh \\
aakar.dhakal@g.bracu.ac.bd}}

\maketitle
\thispagestyle{firstpage} 

\afterpage{\pagestyle{plain}}

\begin{abstract}
 In the era of the internet and smart devices, the detection of malware has become crucial for system security. Malware authors increasingly employ obfuscation techniques to evade advanced security solutions, making it challenging to detect and eliminate threats. Obfuscated malware, adept at hiding itself, poses a significant risk to various platforms, including computers, mobile devices, and IoT devices. Conventional methods like heuristic-based or signature-based systems struggle against this type of malware, as it leaves no discernible traces on the system. In this research, we propose a simple and cost-effective obfuscated malware detection system through memory dump analysis, utilizing diverse machine-learning algorithms. The study focuses on the CIC-MalMem-2022 dataset, designed to simulate real-world scenarios and assess memory-based obfuscated malware detection. We evaluate the effectiveness of machine learning algorithms, such as decision trees, ensemble methods, and neural networks, in detecting obfuscated malware within memory dumps. Our analysis spans multiple malware categories, providing insights into algorithmic strengths and limitations.
By offering a comprehensive assessment of machine learning algorithms for obfuscated malware detection through memory analysis, this paper contributes to ongoing efforts to enhance cybersecurity and fortify digital ecosystems against evolving and sophisticated malware threats. The source code is made open-access for reproducibility and future research endeavours. It can be accessed at \href{https://bit.ly/MalMemCode}{https://bit.ly/MalMemCode}

\end{abstract}

\begin{IEEEkeywords}
System Security, Obfuscated Malware, Malware Categorization, Memory Dump Analysis, Machine Learning, Malware Detection
\end{IEEEkeywords}
\section{Introduction}
The rise of internet connectivity and smart devices has transformed various sectors, but it has also led to an evolving threat landscape, including sophisticated malware targeting interconnected systems. Obfuscated malware, adept at concealing itself, presents a significant challenge to conventional cybersecurity methods. Traditional heuristic-based or signature-based systems struggle to identify such elusive threats, necessitating a shift towards innovative and adaptive detection mechanisms.

This paper explores obfuscated malware detection through multiclass classification, aiming to bridge the gap between evolving threats and advanced detection methods using machine learning. We analyze various algorithms, including decision trees, ensemble methods, support vector machines, and neural networks, to uncover their capabilities and limitations in identifying obfuscated malware.

Acknowledging the significance of class imbalance in real-world datasets, especially in malware detection, we investigate techniques such as undersampling (Edited Nearest Neighbor Rule, Near Miss Rule, Random Undersampling, and All KNN Undersampling) and synthetic data generation using the ADASYN method to address this challenge.

Our research, based on the CIC-MalMem-2022 dataset, simulates real-world scenarios for memory-based obfuscated malware detection. By meticulously analyzing machine learning algorithms and data balancing techniques, we contribute to fortifying cybersecurity against evolving malware threats.

In the following sections, we delve into our dataset, methodologies, and results, aiming to provide valuable insights that can shape the future of malware detection and cybersecurity strategies amidst the challenges posed by obfuscated malware and class imbalance.

\section{Literature Review}

In response to the escalating complexity of malware, researchers have been exploring innovative approaches to strengthen cybersecurity efforts. The development and application of advanced techniques, such as machine learning algorithms and behavioural analysis, have shown promise in augmenting the accuracy and efficiency of malware detection. Furthermore, as malware authors become more adept at evading detection, the focus has shifted toward understanding and countering the methods used for obfuscation.

The study \cite{chen2021malware} discussed the use of static disassembly and machine learning for malware classification, proposing four easy-to-extract and small-scale features for classification. The authors compare their proposed features with detailed behaviour-related features like API sequences and show that the proposed features provide macroscopic information about malware, achieving high accuracy with a smaller feature vector. The paper discusses various approaches to malware analysis combined with machine learning and hand-designed static features. Some approaches mentioned include using strings, registry changes, and API sequences for distinguishing malware variants, extracting 3 grams of byte codes, and using image representations with the K-Nearest Neighbors algorithm. The study also mentions the challenges in malware analysis, such as concept drift and the need for interpretability and explainability in models and features.

According to another study \cite{ahmadi2016novel}, malware analysis involves two main tasks: malware detection and malware classification, with the latter assigning each sample to the correct malware family. Malware classification systems can be divided into two groups: dynamic analysis and static analysis. Dynamic analysis methods capture the behaviour of the program at runtime by monitoring interactions with the operating system, such as analysing API calls and their temporal order or extracting behavioural graphs based on API call parameters. Static analysis methods perform analysis without executing the program, using techniques like pattern detection, bytecode analysis, or disassembling the code to extract information on the program's content. Both dynamic and static analysis rely on the extraction of features for malware detection and classification. This paper focuses on malware classification based on the extraction of static features from PE malware designed for MS Windows systems.

Another paper \cite{you2010malware} explored malware obfuscation techniques, including encrypted, oligomorphic, polymorphic, and metamorphic malwares, which are used to evade antivirus scanners. It provides an overview of the history of malwares developed to defeat signature-based antivirus scanners. The paper discusses the obfuscation techniques commonly used by polymorphic and metamorphic malwares, such as subroutine reordering, instruction substitution, and code transposition. It highlights the limitations of encrypted malwares and the development of oligomorphic and polymorphic malwares to overcome those limitations. The paper also mentions the use of emulation and armouring techniques by antivirus tools to detect and prevent polymorphic malwares.

The paper \cite{kim2018multimodal} proposes a novel framework for Android malware detection using various features and a multimodal deep-learning method. The framework is based on static analysis and aims to distinguish between malware and benign applications. The authors compare the accuracy of their model with other deep neural network models and evaluate the framework's performance in terms of efficiency in model updates, the usefulness of diverse features, and the feature representation method. The proposed framework outperforms a previously proposed method in terms of detection accuracy, precision, recall, F-measure, and overall accuracy. The features generated by the framework effectively capture application characteristics for malware detection. The paper introduces the first application of multimodal deep learning to Android malware detection and provides experimental results to evaluate the performance of the framework.

The current paper \cite{bacci2018impact}demonstrated experimentally that dynamic analysis-based detection performs equally well in evaluating obfuscated and non-obfuscated malware, while static analysis-based detection is more accurate on non-obfuscated samples but is greatly negatively affected by obfuscation. However, this effect can be mitigated by using obfuscated samples in the learning phase.

The paper \cite{aslan2020comprehensive} provides a comprehensive review of malware detection approaches and recent detection methods, aiming to help researchers gain a general understanding of these approaches and their pros and cons. The authors discuss signature-based and heuristic-based detection approaches, which are fast and efficient for known malware but struggle with unknown malware. They also mention behaviour-based, model-checking-based, and cloud-based approaches, which perform well for unknown and complicated malware. Additionally, deep learning-based, mobile devices-based, and IoT-based approaches are emerging to detect both known and unknown malware. However, no single approach can detect all malware in the wild. The literature review section of the paper mentions several related studies. For example, \cite{wagner} proposes a technique to extract malware behaviours from system calls and use a phylogenetic tree to improve the classification. Fukushima et al \cite{fukusima} propose a behaviour-based detection approach for Windows OS, and \cite{chandramohan2013scalable} suggest a bounded feature space behaviour modelling method. Other studies focus on system-centric behaviour models and hardware-enhanced architectures for malware detection.

\section{Methodology}
\subsection{Dataset Description}
The obfuscated malware dataset is a collection of memory dumps from benign and malicious processes, created to evaluate the performance of obfuscated malware detection methods. The dataset contains 58,596 records, with 50\% benign and 50\% malicious. The malicious memory dumps are from three categories of malware: Spyware, Ransomware, and Trojan Horse. The dataset is based on the work \cite{carrier}, who proposed a memory feature engineering approach for detecting obfuscated malware.

The dataset has 55 features and two types of labels, one denoting if they are malicious and the other denoting their malware family. Of the malicious samples,  32.5\% are Trojan Horse malware, 33.67\% are Spyware and 33.8\% are Ransomware samples. 

\begin{figure}[!h]
\includegraphics[scale=0.3]{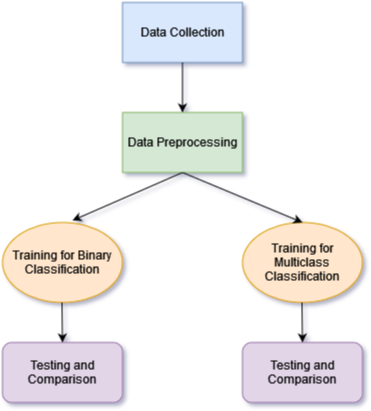}
\centering
\caption{Our work sequence. \label{workflow}}
\end{figure}

After collecting the data, we went through general preprocessing steps and conducted two sets of classification. One to detect malware and the other to classify the malware. Afterwards, we compiled the results to compare. Fig.\ref{workflow} demonstrates our workflow.

\subsection{Data Preprocessing}
At first, we cleaned the data. Some labels had unnecessary information, so we had to trim them. We also standardized the data points for equal treatment of features and better convergence. The dataset was well maintained and clean, so we did not have to perform denoising or further preprocessing steps. For some classifications, we had to encode the labels or oversample or undersample the data points.

\subsection{Model Description}

Here, we provide an in-depth analysis of the used machine learning models, discussing the underlying principles, loss functions, activation functions, mathematical functions, strengths, weaknesses, and limitations.

\subsubsection{Random Forest}

The Random Forest classifier is a powerful ensemble learning method that aims to improve predictive performance and reduce overfitting by aggregating multiple decision trees \cite{breiman2001random}. Each decision tree is built on a bootstrapped subset of the data, and features are sampled randomly to construct a diverse ensemble of trees. The final prediction is determined through a majority vote or weighted average of individual tree predictions.

Random Forest employs the Gini impurity as the default loss function for measuring the quality of splits in decision trees. It is defined as:

\[
I_G(p) = 1 - \sum_{i=1}^{c} p_i^2
\]

Where \( c \) is the number of classes, and \( p_i \) is the probability of a sample being classified as class \( i \).

Random Forest is robust against outliers and noisy data due to the ensemble nature that averages out individual tree errors. Also, it is less prone to overfitting compared to individual decision trees. Additionally, Random Forest provides an estimate of feature importance, aiding in feature selection and interpretation of results.

\subsubsection{Multi-Layer Perceptron (MLP) Classifier}

The Multi-Layer Perceptron (MLP) classifier is a type of artificial neural network that consists of multiple layers of interconnected nodes or neurons \cite{rumelhart1986learning} that process information and pass it to the subsequent layer using activation functions. The architecture typically includes an input layer, one or more hidden layers, and an output layer.

They commonly use the cross-entropy loss function for classification tasks which measures the dissimilarity between predicted probabilities and true class labels. Mathematically, it is defined as:

\[
H(p,q) = - \sum_{i} p(i) \log(q(i))
\]

Where \( p(i) \) is the true probability distribution of class \( i \), and \( q(i) \) is the predicted probability distribution.

They can capture complex patterns in data. The architecture's flexibility allows it to model non-linear relationships effectively. With sufficient data and training, MLPs can achieve high predictive accuracy.

They require careful hyperparameter tuning to prevent overfitting, and their performance heavily depends on the choice of architecture and initialization. 

\subsubsection{k-Nearest Neighbors (KNN) Classifier}

The k-Nearest Neighbors (KNN) classifier is a non-parametric algorithm used for both classification and regression tasks \cite{cover1967nearest}. It makes predictions based on the majority class of the \(k\) nearest data points in the feature space.

KNN is not driven by explicit loss functions or optimization during training. Instead, during inference, it calculates distances between the query data point and all training data points using a distance metric such as Euclidean distance or Manhattan distance. The \(k\) nearest neighbours are then selected based on these distances.
It can handle multi-class problems and adapt well to changes in data distribution. KNN does not assume any underlying data distribution, making it suitable for various data types.
One limitation of KNN is its sensitivity to the choice of \(k\) value and distance metric. 

\subsubsection{XGBoost Classifier}

The XGBoost (Extreme Gradient Boosting) classifier is an ensemble learning algorithm that has gained popularity for its performance and scalability in structured data problems \cite{chen2016xgboost}. It combines the power of gradient boosting with regularization techniques to create a robust and accurate model.
It optimizes a loss function through a series of decision trees. The loss function is often a sum of a data-specific loss (e.g., squared error for regression or log loss for classification) and a regularization term. It minimizes this loss by iteratively adding decision trees and adjusting their weights.
XGBoost excels in handling structured data, producing accurate predictions with fast training times. Its ensemble of shallow decision trees helps capture complex relationships in the data. Regularization techniques, such as L1 and L2 regularization, prevent overfitting and enhance generalization. However, it may not perform as well on text or image data, as it is primarily designed for structured data.

\subsection{Binary Classification}
In this step, our primary target was to detect a malicious memory dump. For this purpose, we have used some simple yet robust machine learning classifiers namely Random Forest Classifier, Multi Level Perceptron and KNN and received outstanding results. Our trained models could detect all the malware accurately and showed great performance in the segregation of malware and safe memory dumps. 

\subsection{Malware Classification}

Here, we tried to separate the malwares from each other based on their different features. Considering the benign samples, the data was highly imbalanced, so we had to go through some further preprocessing before training the detection system. We conducted the classification in three steps:

\begin{itemize}
    \item Classifying in the original format: We ran the data through different ML models and stored the metric scores for comparison. We did not go through any further steps than our usual preprocessing. We passed the same data as we did in binary classification.

    \item Undersampling the majority class: As "Benign" was the majority class in the malware dataset, we undersampled it through various methods, namely Edited Nearest Neighbor Rule\cite{edited}, Near Miss Rule \cite{jiang2018credit}, Random Undersampling and All KNN Undersampling.

    \item Oversampling other minority Classes: We have used the ADASYN \cite{adasyn} method to generate synthetic data points for all minority classes.
\end{itemize}
After applying these techniques, our training data(80\% of the total data) size is shown in TABLE \ref{size}:
\begin{table}[!ht]
    \centering
    \caption{Training size of the original, undersampled and oversampled dataset.}
    \begin{tabular}{ll}
    \hline
        \textbf{Data Type} & \textbf{Training Size} \\ \hline
        Original & 46876 \\ \midrule
        Edited Nearest Neighbor & 36157 \\ 
        Near Miss & 30356 \\
        All KNN & 41159 \\
        Random Undersampling & 30356 \\ \midrule
        ADASYN & 117193 \\ \hline
    \end{tabular}
    \label{size}
\end{table}
Afterwards, we used different machine learning algorithms to train our detection models and recorded their performance.

\section{Results and Discussion}
From our experiments, we have achieved outstanding results on our malware detection system.
\subsection{Binary Classification}
Our trained model achieved 99.99\% accuracy on the test set, detecting all the malware correctly. The result is shown in Fig.\ref{bin}
\begin{figure}[!h]
\includegraphics[scale = 0.3]{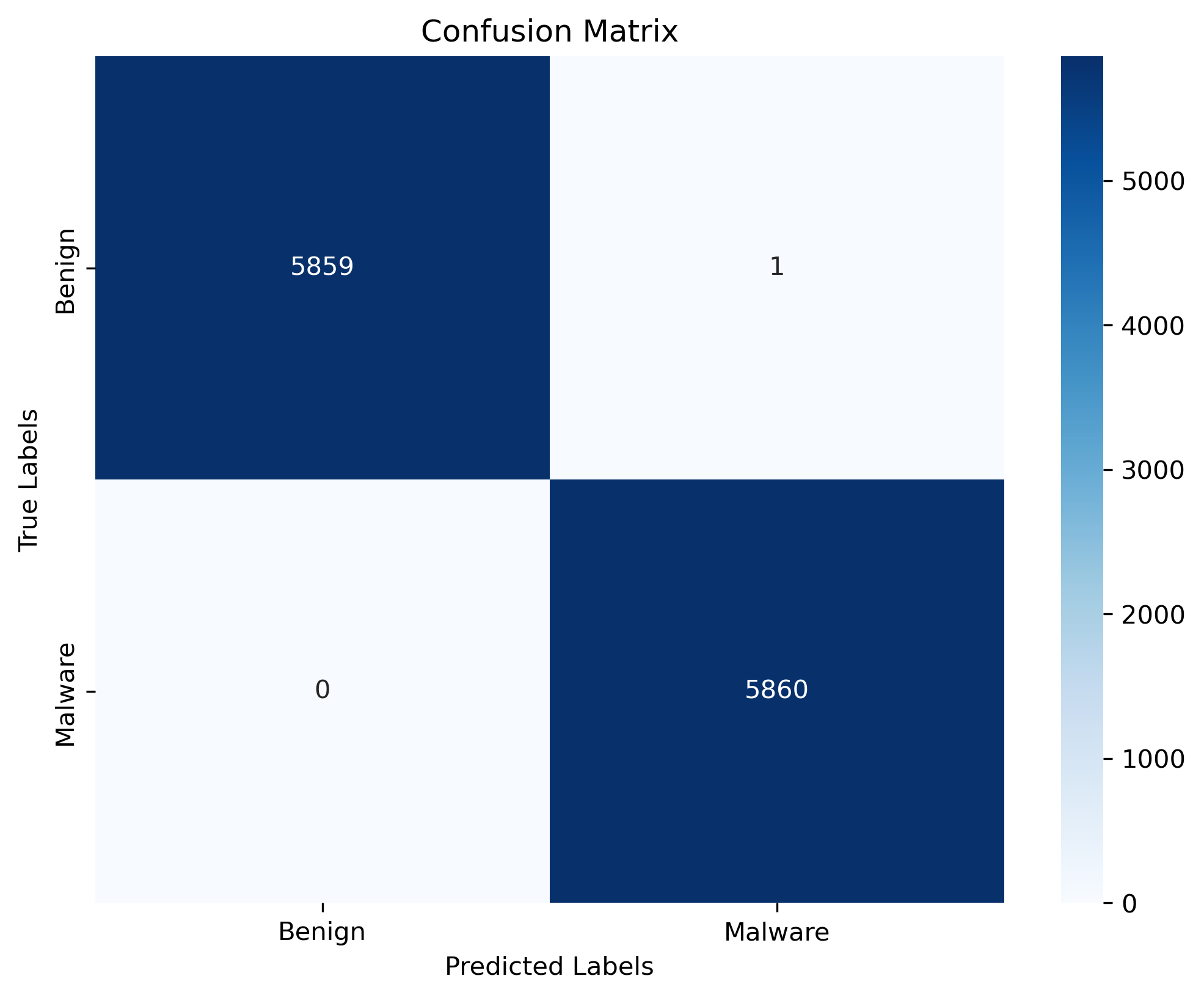}
\centering
\caption{Binary classification using Random Forest Classifier. \label{bin}}
\end{figure}
However, all the models performed very well in the detection of potential malware. The results are tabulated in TABLE \ref{bin}

\begin{table}[!ht]
    \centering
    \caption{Binary Classification Performance}
    \begin{tabular}{|l|l|l|l|l|}
    \hline
        \textbf{Model Name} & \textbf{Accuracy} & \textbf{Precision} & \textbf{Recall} & \textbf{F1 Score} \\ \hline
        Random Forest & 0.9999 & 0.9999 & 0.9999 & 0.9999 \\ \hline
        MLP & 0.9999 & 0.9999 & 0.9999 & 0.9999 \\ \hline
        KNN & 0.9991 & 0.9991 & 0.9991 & 0.9991 \\ \hline
        XGBoost & 0.9999 & 0.9999 & 0.9999 & 0.9999 \\ \hline
    \end{tabular}
    
    \label{bin}
\end{table}

\subsection{Malware Classification}
As the dataset is highly imbalanced, we conducted this part in 3 steps. First, we conducted the experiment on the original dataset, then undersampled the majority class and later oversampled the minority classes.
\subsubsection{Classification on Original Data}
Here, we ran the untouched data through our chosen algorithms and achieved moderate results. Although the metrics are not as impressive as the binary classification, it is mentionable that, no malware was classified safe, rather, different malwares were classified wrong. Our result is tabulated in TABLE \ref{or}.
\begin{table}[!ht]
    \centering
    \caption{Classification on original data}
    \begin{tabular}{|l|l|l|l|l|}
    \hline
        \textbf{Model Name} & \textbf{Accuracy} & \textbf{Precision} & \textbf{Recall} & \textbf{F1 Score} \\ \hline
        Random Forest & 0.8721 & 0.8719 & 0.8721 & 0.8720 \\ \hline
        MLP & 0.7766 & 0.7765 & 0.7766 & 0.7765 \\ \hline
        KNN & 0.8171 & 0.8194 & 0.8171 & 0.8172 \\ \hline
        XGBoost & 0.8815 & 0.8812 & 0.8815 & 0.8813 \\ \hline
    \end{tabular}
    
    \label{or}
\end{table}
From the results, it is seen that the XGBoost classifier performed the best in the detection and classification of malware.

\subsubsection{Undersampling Majority Class}
We have used four types of undersampling methods and trained our models on all of them. We got different performance metrics for different undersampling methods. No single method could dominate the scores.  However, Random Undersampling and Near Miss approaches performed better than the other two methods. These results are tabulated in TABLE \ref{under}. 
\begin{table*}[!ht]
    \centering
    \caption{Classification on undersampled data}
    \begin{tabular}{|l|l|l|l|l|l|}
    \hline
        \textbf{Model Name} & \textbf{Undersampling Technique} & \textbf{Accuracy} & \textbf{Precision} & \textbf{Recall} & \textbf{F1 Score} \\ \hline
        Random Forest & Random Undersampling & 0.8716 & 0.8716 & 0.8716 & 0.8716 \\ \hline
        MLP & Near Miss & 0.7775 & 0.7822 & 0.7775 & 0.7783 \\ \hline
        KNN & Near Miss & 0.8120 & 0.8147 & 0.8120 & 0.8124 \\ \hline
        XGBoost & Random Undersampling & 0.8784 & 0.8783 & 0.8784 & 0.8783 \\ \hline
    \end{tabular}
    
    \label{under}
\end{table*}
From the results, we can see, that the XGBoost Classifier also performed better in this case while the Random Forest Classifier was really close. In this approach too, no malware was labeled safe during detection.

\subsubsection{Oversampling Minority Classes}

Among the popular oversampling methods, we choose ADASYN(Adaptive Synthetic Sampling). It is a data augmentation technique primarily used in imbalanced classification tasks. After applying ADASYN to all the minority classes separately, we balanced the dataset and applied our chosen classification algorithms. We got our best results with this approach. The findings are tabulated in TABLE \ref{over}
\begin{table}[!ht]
    \centering
    \caption{Performance on Oversampled Data}
    \begin{tabular}{|l|l|l|l|l|}
    \hline
        \textbf{Model Name} & \textbf{RF} & \textbf{MLP} & \textbf{KNN} & \textbf{XGBoost} \\ \hline
        Accuracy & 0.9395 & 0.8211 & 0.9121 & 0.9427 \\ \hline
        Precision & 0.9395 & 0.8211 & 0.9129 & 0.9427 \\ \hline
        Recall & 0.9395 & 0.8211 & 0.9121 & 0.9427 \\ \hline
        F1 Score & 0.9395 & 0.8211 & 0.9122 & 0.9427 \\ \hline
    \end{tabular}
    
    \label{over}
\end{table}

\begin{figure}[!h]
\includegraphics[scale = 0.3]{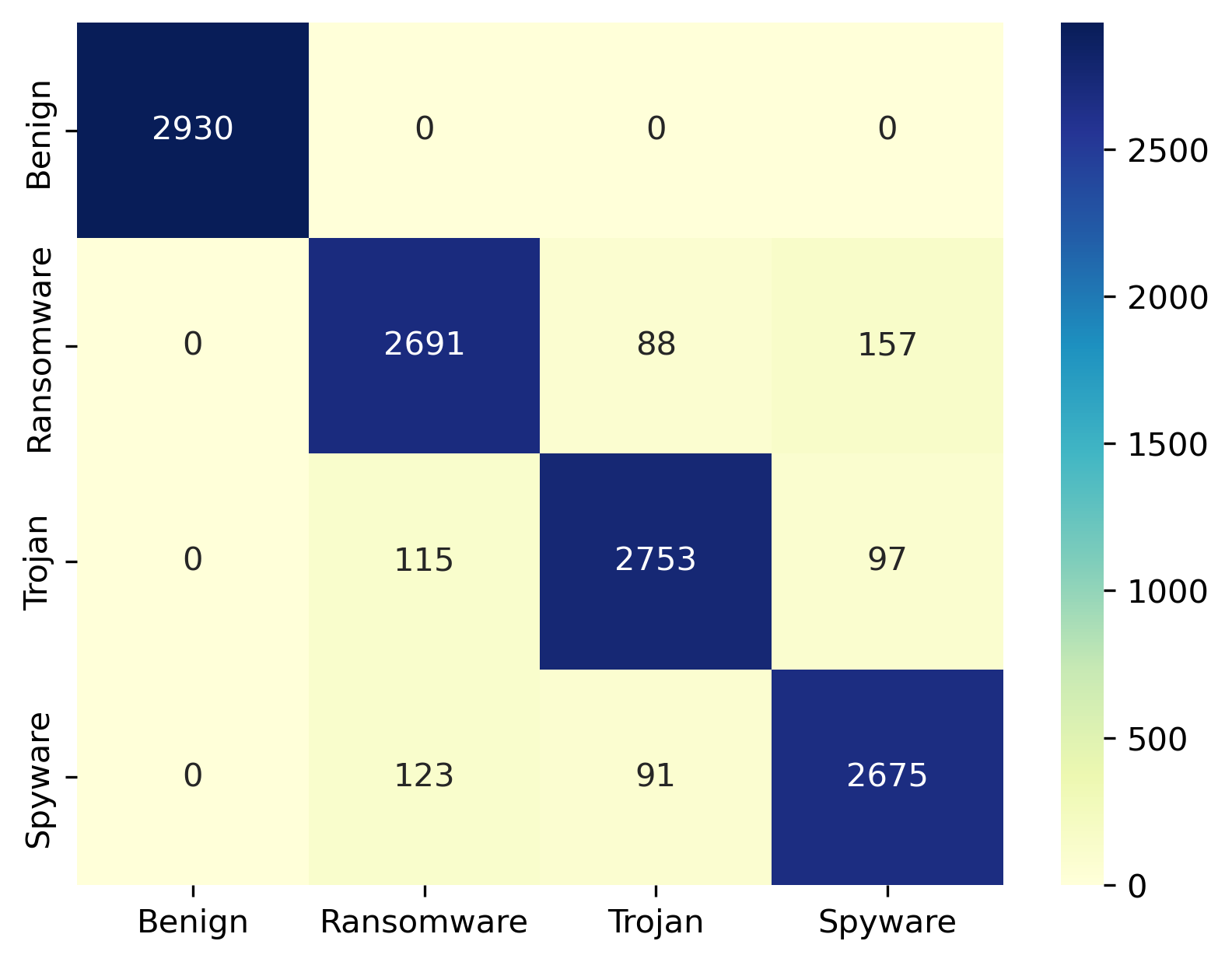}
\centering
\caption{XGBoost performance in detection. \label{xgb}}
\end{figure}

Here also, XGBoost outperformed the other classifiers and provided the best predictions. The detection is shown in the Fig.\ref{xgb}

Therefore, we see that our malware detection models are well-performing and robust. It can perfectly detect any potential malware through memory dump analysis as we conduct binary classification.
In classifying the malware, among the explored approaches, the application of ADASYN emerged as the most promising solution. By systematically addressing the class imbalance through synthetic data generation, we achieved superior results compared to both the original format classification and the undersampling techniques. The outcomes of our experiments underscore the importance of tailored strategies for handling class imbalance and reaffirm the potential of advanced techniques like ADASYN in enhancing multiclass classification accuracy.

\section{Conclusion and Future Work}

In conclusion, our research addresses the rising threat of obfuscated malware in connected devices and the internet landscape. Through memory dump analysis and diverse machine learning algorithms, we've explored effective detection strategies and illuminated their strengths and limitations using the CIC-MalMem-2022 dataset. Emphasizing the synergy between machine learning and traditional security methods, our work underscores the need for a comprehensive defense strategy in the dynamic cybersecurity realm. While acknowledging the ever-evolving malware landscape, our research lays the groundwork for future endeavours, advocating continuous adaptation. Future efforts should focus on refining algorithms, exploring new data sources, and fostering interdisciplinary collaboration. We envision research on hybrid approaches, combining machine learning and signature-based methods, and studying the impact of adversarial attacks and explainable AI to enhance detection system robustness and transparency. In summary, our study provides valuable insights for resilient cybersecurity solutions, addressing the challenges of obfuscated malware and advancing detection capabilities to safeguard digital ecosystems against emerging threats.

\bibliographystyle{IEEEtran}
\bibliography{sample-base}
\end{document}